\documentclass[doublecol]{epl2}

\usepackage{graphicx}
\usepackage{amsmath}
\usepackage{color}
\usepackage{hyperref}

\title{Hall effect and Fermi surface reconstruction via electron pockets in the high-$T_c$ cuprates.}
\shorttitle{Hall effect and Fermi surface reconstruction in the high-$T_c$ cuprates.} 

\author{J.G. Storey\inst{1}}
\shortauthor{J.G. Storey}

\institute{                    
  \inst{1} Robinson Research Institute, Victoria University - 
P.O. Box 600, Wellington, New Zealand
}


\abstract{
The mechanism by which the Fermi surface of high-$T_c$ cuprates undergoes a dramatic change from a large hole-like barrel to small arcs or pockets on entering the pseudogap phase remains a question of fundamental importance.
Here we calculate the normal-state Hall coefficient from the resonating-valence-bond spin-liquid model developed by Yang, Rice and Zhang. In this model, reconstruction of the Fermi surface occurs via an intermediate regime where the Fermi surface consists of both hole- and electron-like pockets. We find that the doping $(x)$ dependence of the Hall number transitions from $1+x$ to $x$ over this narrow doping range. At low temperatures, a switch from a downturn to an upturn in the Hall coefficient signals the departure of the electron-like pockets from the Fermi surface.
}

\begin{document}

\maketitle

As we approach the thirtieth anniversary of the discovery of cuprate high-temperature superconductivity a growing breed of high-field experiments is pushing back the veil of superconductivity to lower temperatures, providing opportunities to answer fundamental questions about the underlying normal state. Key among these are the following, how exactly does the Fermi surface transition from a large hole-like barrel in overdoped samples\cite{PLATEPRL,VIGNOLLE,Fujita09052014}, to the small arcs or pockets\cite{Shen11022005,Jaudet2009354,Fujita09052014} associated with the pseudogap phase? Are electron-like pockets part of this process? Where does the onset of the pseudogap occur?  Answering these questions will go a long way towards explaining the origin of the enigmatic pseudogap, which adversely affects superconductivity over a wide range of temperature and doping\cite{TALLON2}.

\begin{figure}
\centering
\includegraphics[width=\linewidth]{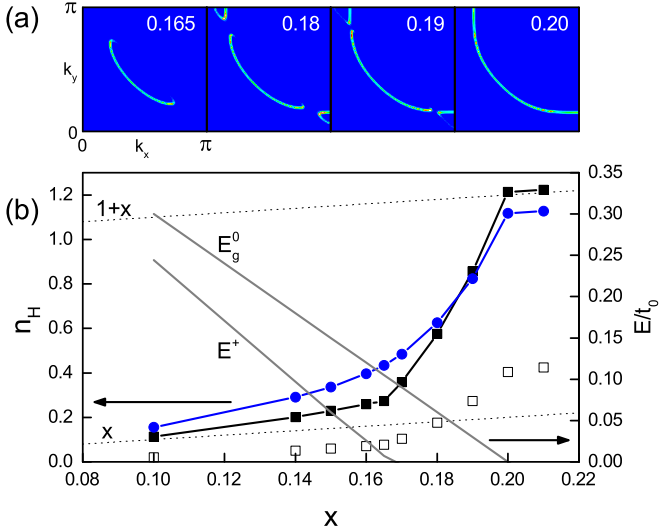}
\caption{
(Color online) (a) Spectral weight at the Fermi level calculated from the YRZ model for the values of $x$ shown in each panel. (b) Zero-temperature Hall number vs doping for scattering rates given by eq.~\ref{eq:GAMMA2} (filled squares), and $\Gamma^\pm=0.01t_0$ (circles). Including the $g_t(x)$ prefactor from eq.~\ref{eq:GYRZ2} in the spectral function (eq.~\ref{eq:AKW}) results in values that are too small (open squares). The pseudogap magnitude $E_g^0$ and the energy of the bottom of the upper branch $E^+$ are shown by the grey lines.
} 
\label{NHFIG}
\end{figure}

In this work we calculate the normal-state Hall coefficient from the resonating-valence-bond spin-liquid ansatz of Yang, Rice and Zhang (YRZ)\cite{YRZ}, with the hope of inspiring future experiments to address these issues.
Our focus here is the pseudogap onset near a doping $x=0.19$\cite{OURWORK1}, not the charge-density-wave (CDW) reconstruction centered around 0.125\cite{LALIBERTE} which seems to be an additional or subsidiary effect\cite{Comin24012014,daSilvaNeto24012014,WU}.
In the YRZ model a pseudogap term in the self-energy reconstructs the energy-momentum dispersion into two branches. The lower branch produces a hole-like pocket at the nodal regions of the Brillouin zone. When the pseudogap is small, parts of the upper branch cross below the Fermi level resulting in electron pockets near the antinodes (see fig.~\ref{NHFIG}(a)).
We have chosen this model because it successfully describes experimental data from a wide range of techniques\cite{RICE,ASHBY,STOREYEPL2012}. While there exists angle-resolved photoemission spectroscopy (ARPES) data consistent with the shape of the YRZ hole pockets\cite{YANG2}, direct evidence for the electron pockets is still lacking. Tentative indirect evidence, however, may be found in an observed undulation in the temperature-dependence of the thermoelectric power\cite{STOREYTEP}.

For detailed descriptions of the YRZ model we refer the reader to refs.~\cite{YRZ,SCHACHINGER,BORNE1}. The equations used in this work are presented below.
In the normal state the coherent part of the electron Green's function is given by
\begin{equation}
G(\textbf{k},\omega,x)=\frac{g_t(x)}{\omega-\xi_\textbf{k}-\frac{E_g^2(\textbf{k})}{\omega+\xi_\textbf{k}^0}}
\label{eq:GYRZ}
\end{equation}
where $\xi_\textbf{k}=-2t(x)(\cos k_x+\cos k_y)
-4t^\prime(x)\cos k_x\cos k_y
-2t^{\prime\prime}(x)(\cos 2k_x+\cos 2k_y)-\mu_p(x)$ is the tight-binding energy-momentum dispersion, $\xi_\textbf{k}^0=-2t(x)(\cos k_x+\cos k_y )$ is the nearest-neighbour term, and $E_g(\textbf{k})=[E_g^0(x)/2](\cos k_x-\cos k_y)$
is the pseudogap with $E_g^0(x)=3t_0(0.2-x)$ for $x\leq 0.2$, while for $x>0.2$ $E_g^0(x)=0$. The chemical potential $\mu_p(x)$ is chosen according to the Luttinger sum rule.
The doping-dependent coefficients are given by $t(x)=g_t(x)t_0+(3/8)g_s(x)J\chi$, $t^\prime(x)=g_t(x)t_0^\prime$ and $t^{\prime\prime}(x)=g_t(x)t_0^{\prime\prime}$, where $g_t(x)=2x/(1+x)$ and $g_s(x)=4/(1+x)^2$ are the Gutzwiller factors. The bare parameters $t^\prime/t_0=-0.3$, $t^{\prime\prime}/t_0=0.2$, $J/t_0=1/3$ and $\chi=0.338$ are the same as used previously\cite{YRZ}.
Equation~\ref{eq:GYRZ} can be re-written as
\begin{equation}
G(\textbf{k},\omega,x)=\sum_{\alpha=\pm}{\frac{g_t(x)W_\textbf{k}^\alpha(x)}{\omega-E_\textbf{k}^\alpha(x)}}
\label{eq:GYRZ2}
\end{equation}
where the energy-momentum dispersion is reconstructed by the pseudogap into upper and lower branches
\begin{equation}
E_\textbf{k}^\pm=\frac{1}{2}(\xi_\textbf{k}-\xi_\textbf{k}^0)\pm\sqrt{\left(\frac{\xi_\textbf{k}+\xi_\textbf{k}^0}{2}\right)^2+E_g^2(\textbf{k})}
\label{eq:EK}
\end{equation}
that are weighted by
\begin{equation}
W_\textbf{k}^\pm=\frac{1}{2}\left[1\pm\frac{(\xi_\textbf{k}+\xi_\textbf{k}^0)/2}{\sqrt{[(\xi_\textbf{k}+\xi_\textbf{k}^0)/2]^2+E_g^2(\textbf{k})}}\right]
\label{eq:WK}
\end{equation}
The spectral function is
\begin{equation}
A_\textbf{k}^\pm(\omega)=\frac{1}{\pi}\frac{W_\textbf{k}^\pm\Gamma^\pm}{(\omega-E_\textbf{k}^\pm)^2+(\Gamma^\pm)^2}
\label{eq:AKW}
\end{equation}
$\Gamma^\pm$ is the scattering rate or inverse lifetime of the quasiparticles. The weight functions, $W_\textbf{k}^\pm$, give the impression of arcs rather than pockets in plots of the spectral function at the Fermi energy, $\omega=0$, as demonstrated in fig.~\ref{NHFIG}(a). Note that we have dropped the $g_t(x)$ prefactor in eq.~\ref{eq:AKW}, effectively taking the Green's function to be completely coherent\cite{BORNE1}. As will be seen later in fig.~\ref{NHFIG}(b) retaining the prefactor produces Hall number values that are too small. $g_t(x)$ is still present in the tight-binding coefficients $t(x)$, $t^\prime(x)$ and $t^{\prime\prime}(x)$. 

The density of states is
\begin{equation}
N^\pm(\omega)=\sum_\textbf{k}{A^\pm_\textbf{k}(\omega)}
\label{eq:DOS}
\end{equation}

Finally, the Hall coefficient is calculated from\cite{VORUGANTI}
\begin{equation}
R_H = \frac{\sigma_{xy}^-+\sigma_{xy}^+}{(\sigma_{xx}^-+\sigma_{xx}^+)(\sigma_{yy}^-+\sigma_{yy}^+)}
\label{RHEQ}
\end{equation}
in terms of the conductivities
\begin{equation}
\begin{split}
& \sigma_{xy}^\pm = \frac{4\pi^2e^3}{3V}\int d\omega\left(\frac{\partial f(\omega)}{\partial\omega}\right)\times\\&\frac{1}{N}\sum_\textbf{k}v_x^\pm(\textbf{k})\biggl(v_x^\pm(\textbf{k})\frac{\partial v_y^\pm(\textbf{k})}{\partial k_y}-v_y^\pm(\textbf{k})\frac{\partial v_y^\pm(\textbf{k})}{\partial k_x}\biggr)[A_\textbf{k}^\pm(\omega)]^3
\end{split}
\label{eq:ASIGMAXYZ}
\end{equation}
and
\begin{equation}
\begin{split}
\sigma_{xx}^\pm & = \sigma_{yy}^\pm \\& = \frac{2\pi e^2}{V}\int d\omega\left(-\frac{\partial f(\omega)}{\partial\omega}\right)\frac{1}{N}\sum_\textbf{k}{[v_x^\pm(\textbf{k})]^2[A_\textbf{k}^\pm(\omega)]^2}
\end{split}
\label{eq:ASIGMAXX}
\end{equation}
where $v_\alpha^\pm(\textbf{k})=\partial E_\textbf{k}^\pm/\partial k_\alpha$, $f$ is the Fermi function and $V$ is the normalisation volume. Note that the velocities must be calculated from each of the reconstructed branches $E_\textbf{k}^\pm$, rather than $\xi_\textbf{k}$. 

In the following calculations the scattering rate is taken to be
\begin{equation}
\Gamma^\pm(\textbf{k})=\Gamma_0+C\sqrt{[v_x^\pm(\textbf{k})]^2+[v_y^\pm(\textbf{k})]^2}
\label{eq:GAMMA2}
\end{equation}
where $\Gamma_0$ is a small value to prevent $\Gamma^\pm$ going to zero at the saddle points at $(\pm\pi,0)$ and $(0,\pm\pi)$, and $C$ is an adjustable constant initially set to $0.01t_0$. A value of $0.1t_0$ was also investigated, but this did not alter the conclusions drawn here. A scattering rate of this form corresponds to a constant mean free path, which has been found to account for the thermopower of optimal to overdoped hole-doped cuprates\cite{KONDOTEP,STOREYTEP}. It also reproduces the increased lifetime near the saddle points observed by ARPES in overdoped samples\cite{PLATEPRL,YANGK}. 

\begin{figure}
\centering
\includegraphics[width=\linewidth]{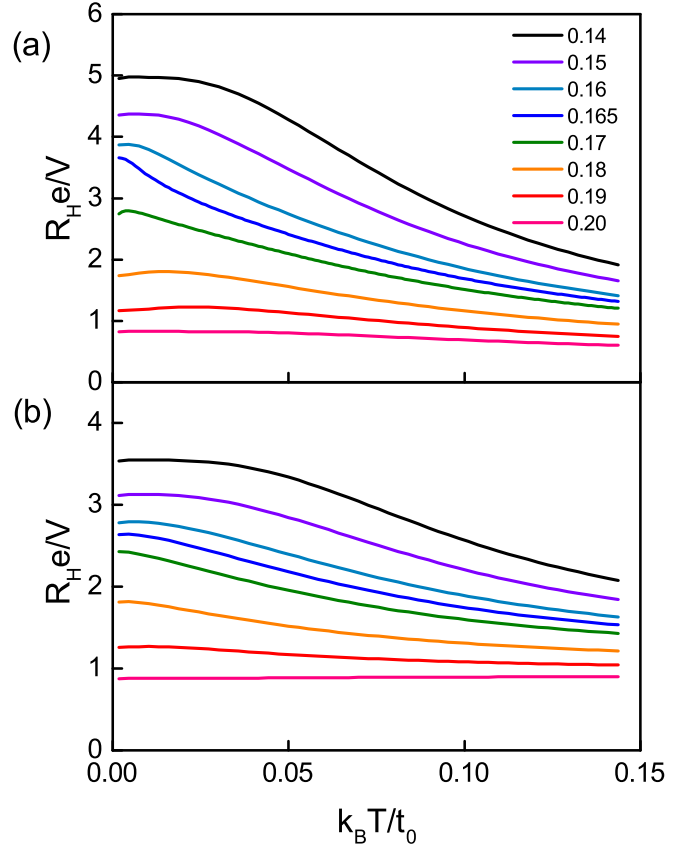}
\caption{
(Color online) Normal-state Hall coefficient vs temperature for dopings $x=0.14$ to 0.20, with scattering rates given by (a) eq.~\ref{eq:GAMMA2}, and (b) $\Gamma^\pm=0.01t_0$.
} 
\label{RHVSTFIG}
\end{figure}

\begin{figure}
\centering
\includegraphics[width=\linewidth]{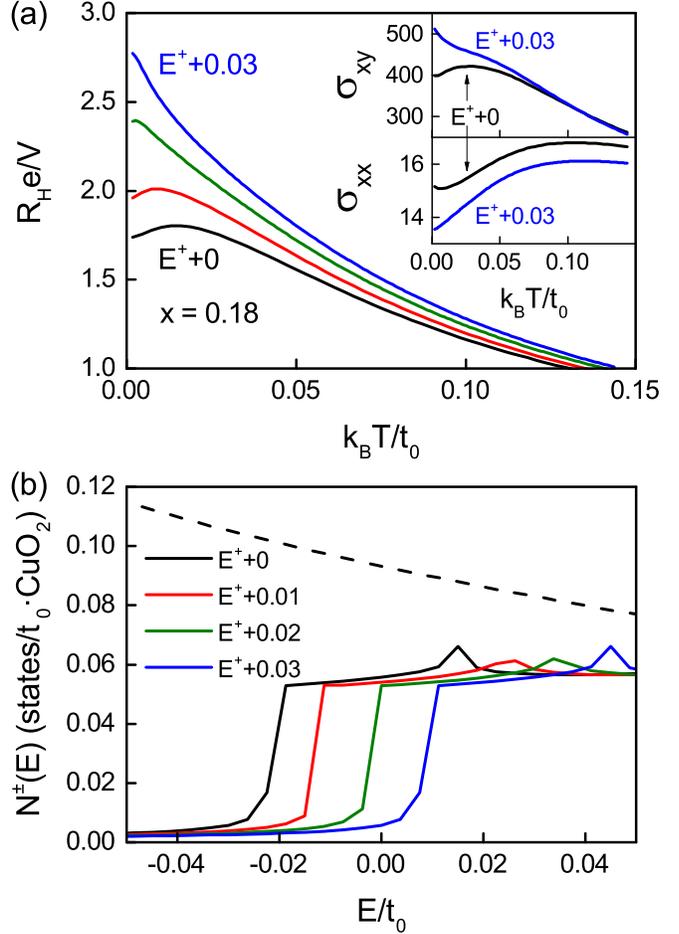}
\caption{
(Color online) (a) Hall coefficient vs temperature for $x=0.18$, with the upper branch, $E_\textbf{k}^+$ progressively shifted to higher energies. Insets: Conductivities $\sigma_{xy}(T)$ in $e^3/Vt_0^2$ units, and $\sigma_{xx}(T)$ in $e^2/Vt_0$ units, corresponding to zero and $0.03t_0$ shifts of the upper branch. (b) Partial density of states of the lower (dashed line) and upper (solid lines) branches corresponding to the curves in (a).
} 
\label{BANDPUSHFIG}
\end{figure}

The temperature dependence of the Hall coefficient, $R_H(T)$ in this low scattering regime, is plotted in fig.~\ref{RHVSTFIG}(a) for several dopings $x$ in the range 0.14 to 0.20. The salient feature of this plot is a progression from a low-temperature downturn for $x>0.17$ to an upturn for $x\leq 0.17$.
As will be shown below, this coincides with the lifting of the electron pockets from the Fermi surface. 
Figure~\ref{NHFIG}(a) shows the spectral function at the Fermi level in the first Brillouin zone for selected dopings, highlighting the evolution from a large hole-like Fermi surface to small hole-like Fermi pockets with the intermediate presence of electron pockets. The back side of the nodal pocket, invisible due to the small value of $W_\textbf{k}^-$ there, runs along the zone diagonal $(0,\pi)$ to $(\pi,0)$.
The doping dependence of the zero-temperature Hall number $n_H = V/eR_H$ is shown in fig.~\ref{NHFIG}(b) (filled squares). Above $x=0.2$, $n_H$ is proportional to $1+x$ as expected. The opening of the pseudogap causes a rapid drop in $n_H(x)$ below $x=0.2$. The drop occurs over the range $0.165<x<0.2$ where the Fermi surface is comprised of a nodal hole-like pocket and antinodal electron-like pockets, as indicated by the plots of the spectral function in this region (fig.~\ref{NHFIG}(a)). Although we have taken values from the original YRZ paper\cite{YRZ}, both the critical doping at which the pseudogap opens and $E_g^0(x)$ could be adjusted to tune the location and slope of this intermediate region. Once the upper YRZ-band electron pockets lift above the Fermi level, near $x=0.165$, $n_H$ decreases more gradually tending towards $n_H=x$. (An exact match with $n_H=x$ would be expected in the case of a circular Fermi surface.) Below this doping the pseudogap spans the Fermi level.
To verify that this is the case we plot the Hall coefficient for $x=0.18$ and then progressively shift the upper YRZ band to higher energies, with everything else kept equal (fig.~\ref{BANDPUSHFIG}(a)). The respective densities of states for the upper and lower bands are shown in fig.~\ref{BANDPUSHFIG}(b). The traversal of the edge of the upper band above the Fermi level noticeably alters the low-temperature behaviour of both $\sigma_{xy}(T)$ and $\sigma_{xx}(T)$ (see fig.~\ref{BANDPUSHFIG}(b) insets), resulting in $R_H(T)$ switching from a downturn to an upturn.

\begin{figure}
\centering
\includegraphics[width=\linewidth]{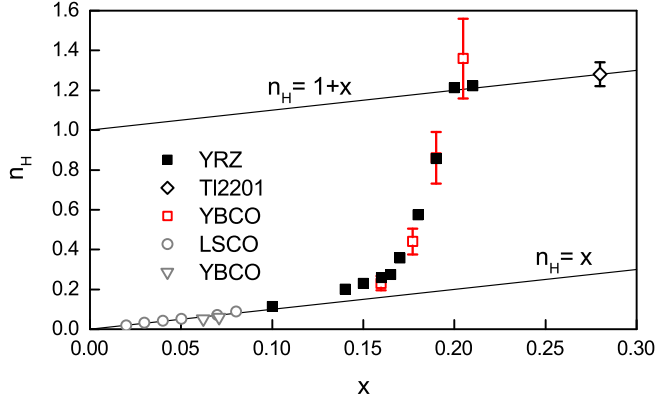}
\caption{
(Color online) Comparison of the Hall number calculated from the YRZ model and eq.~\ref{eq:GAMMA2} (filled squares) with experimental values from Badoux \textit{et al}.\cite{BADOUX} and references therein (open symbols).
} 
\label{EXPDATAFIG}
\end{figure}

These calculations suggest measurements of the normal-state Hall coefficient to low temperatures as a means of testing for the presence of electron pockets near the onset of the pseudogap. Thermoelectric power measurements could be conducted in tandem to look for the predicted undulation\cite{STOREYTEP}. Such experiments would take advantage of the increasingly high magnetic fields available to suppress the superconducting state. However, we caution that high-field studies while exposing the normal state at low temperatures could result in additional Fermi surface reconstruction due to the onset of long-range CDW ordering\cite{WU2011,LEBOEUF}.
It is worth mentioning that the observation of this effect depends on the scattering rate having a momentum dependence similar to eq.~\ref{eq:GAMMA2}, as well as probing to sufficiently low temperatures. For example, the upturn in $R_H(T)$ becomes flattened when an isotropic scattering rate given by $\Gamma^\pm=0.01t_0$ is assumed, as seen fig.~\ref{RHVSTFIG}(b). The drop in $n_H(x)$ is still evident, but it levels out more smoothly, as shown in fig.~\ref{NHFIG}(b) (circles).
The doping range over which the electron pockets are present is reasonably narrow, requiring a set of samples with suitably small doping increments. 
Our results are very consistent with recent Hall measurements on YBa$_2$Cu$_3$O$_y$ at high field\cite{BADOUX}, see for example a comparison of the Hall number in fig.~\ref{EXPDATAFIG}. Ideally a more thorough study should be made as a function of $T$ and $x$ in suitable cuprates, if possible several.

\begin{figure}
\centering
\includegraphics[width=\linewidth]{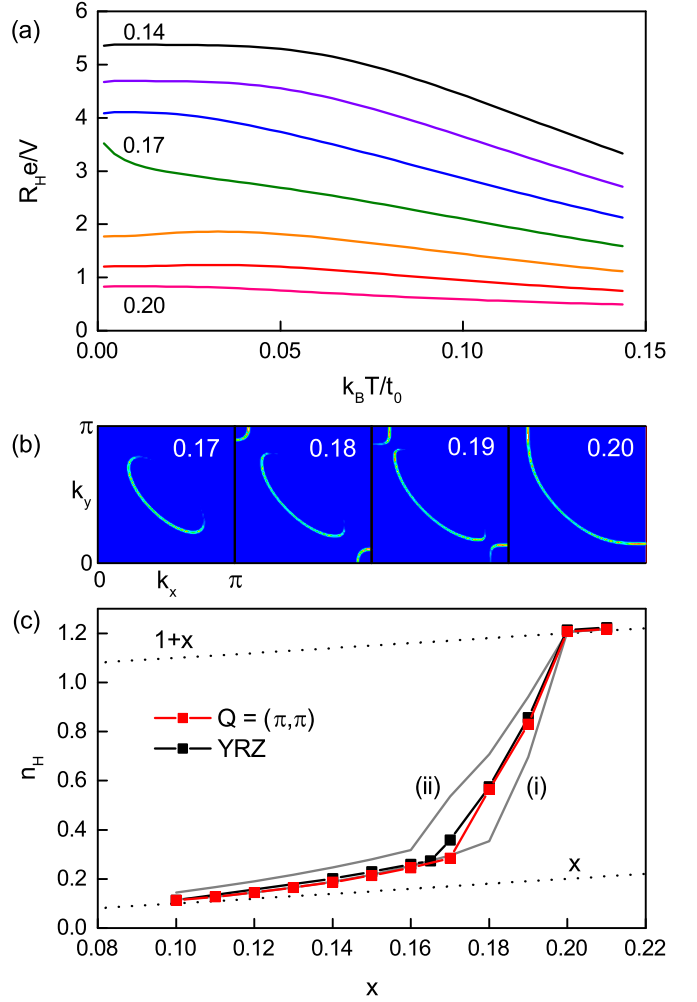}
\caption{
(Color online) (a) Normal-state Hall coefficient for $x$=0.14 to 0.2 calculated from a $\textbf{Q}=(\pi,\pi)$ Fermi surface reconstruction, assuming tight binding coefficients fixed to their values at $x$=0.15 and an isotropic pseudogap given by $E_g(x) = 6t_0(0.2-x)$ for $x<0.2$. (b) Spectral weight at the Fermi level for selected dopings. (c) Hall number corresponding to the curves in (a) (red squares). Curve (i) shows the effect of changing the tight binding coefficients to their values at $x$=0.1. Curve (ii) shows the further effect of halving the pseudogap magnitude to $E_g(x) = 3t_0(0.2-x)$. YRZ-model values for comparison are also shown (black squares).
}
\label{AFFIG}
\end{figure}

As a final point of discussion, it is worth pointing out that similar results are produced by the antiferromagnetic $\textbf{Q}=(\pi,\pi)$ zone-folding reconstruction model\cite{CHUBUKOV}. This model also features antinodal electron pockets and nodal hole pockets with the difference that they are symmetric about the zone diagonal $(\pi,0)$ to $(0,\pi)$. The equations for this scenario are obtained by replacing $\xi_\textbf{k}^0$ by $-\xi_\textbf{k+Q}$ in eqs.~\ref{eq:EK} and \ref{eq:WK}. We assume a scattering rate given by eq.~\ref{eq:GAMMA2}, an isotropic pseudogap $E_g(x)$, and fix the tight binding coefficients to a particular value of $x$. Results are shown in fig.~\ref{AFFIG}. By taking $t(0.15)$, $t^\prime(0.15)$ and $t^{\prime\prime}(0.15)$ with $E_g(x)=6t_0(0.2-x)$ the Hall number comes out almost identical to the YRZ model (fig.~\ref{AFFIG}(c)). For these parameters the temperature dependence of $R_H$ is more gradual than that of YRZ because the pseudogap magnitude is twice as large, but the upturn is still present when the electron pockets rise above the Fermi level. The slope of the transition in the Hall number is sensitive to changes in the tight binding coefficients as well as the pseudogap magnitude (fig.~\ref{AFFIG}(c)). In the hole-pocket regime, the Hall number gets closer to $n_H=x$ for larger values of $E_g(x)$ due to the pockets becoming more circular. This is also the case in the YRZ model.

In conclusion, we have calculated the normal-state Hall coefficient of the high-$T_c$ cuprates assuming a Fermi surface reconstruction model for the pseudogap proposed by Yang, Rice and Zhang\cite{YRZ}. A switch from a downturn to an upturn at low temperatures signals the lifting of antinodal electron-like pockets from the Fermi surface. We propose that a search for this effect be carried out in order to help identify both the mechanism of the reconstruction from a large to a small Fermi surface, and possibly the origin of the pseudogap itself.

\begin{acknowledgments}
This work was supported by the Marsden Fund Council from Government funding, administered by the Royal Society of New Zealand. The author acknowledges helpful discussions with J.L. Tallon who suggested undertaking this study.
\end{acknowledgments}


\end{document}